\documentclass[11pt]{article}
\usepackage{graphicx}

\title{Entropy Generation in Computation and \\
the Second Law of Thermodynamics
}

\author{Shunya Ishioka\\
Department of Information Science, Kanagawa University\\
Hiratsuka, Kanagawa 259-1293, Japan\\
\vspace*{3mm}
e-mail : ishioka@educ.info.kanagawa-u.ac.jp \\
Nobuko Fuchikami\\
Department of Physics, Tokyo Metropolitan
University\\
Minami-Ohsawa, Hachioji, Tokyo 192-0397, Japan\\
e-mail : fuchi@phys.metro-u.ac.jp
}

\date{31 January 1999}

\begin{document}
\maketitle

\begin{abstract}
Landauer discussed the minimum energy necessary for 
computation and stated that
 erasure of information is accompanied by {\it heat} generation 
to the amount of $kT \ln2/$bit.
Modifying the above  statement, we claim that
erasure of information is accompanied by {\it entropy} generation
$k \ln2/$bit.   
Some new concepts will be introduced in the field of thermodynamics that are 
implicitly included in our statement.
The new concepts that we will introduce are 
``partitioned state'' , which corresponds to frozen state such as in 
ice, 
``partitioning process'' and ``unifying process''.
Developing our statement, i.e., our thermodynamics of computation,
we will point out that the so-called ``residual entropy'' does 
not exist in 
the partitioned state. We then argue that a partioning process is 
an entropy decreasing process. Finally we reconsider the second law of
thermodynamics especially when computational processes are involved.
\end{abstract}

\renewcommand{\thesection}{\S\arabic{section}}
\section{Introduction}

\hspace*{5mm}
About forty years ago Landauer discussed the minimum energy necessary for 
computation\cite{l61}. His conclusion is that 

\vspace*{3mm}
 [ L ] erasure of information is accompanied by {\it heat} generation 
to the amount of $kT \ln2/$bit.

\vspace{3mm}
\noindent
It seems that his theory has been accepted widely\cite{b97,f97}. 
In our opinion, however, a more precise expression must be that

\vspace{3mm}
 [ IF ] erasure of information is accompanied by {\it entropy} generation
$k \ln2/$bit.

\vspace{3mm}
The aim of the present paper is not to compare the above two statements 
but to point out some new concepts in the field of thermodynamics which are 
implicitly included in statement [ IF ]. 
The new concepts that we will introduce are 
``partitioned state'' (, which corresponds to frozen state such as in 
ice), 
``partitioning process'' and ``unifying process''.

We first explain [ IF ], namely, our thermodynamics of computation.
It will be pointed out that the so-called ``residual entropy'' does 
not exist in 
the partitioned state. We then argue that a partioning process is 
an entropy decreasing process. Finally we reconsider the second law of
thermodynamics.

\section{Thermodynamics of computation}

\hspace*{5mm}
A hardware with one-bit memory corresponds to
a bistable physical system(, for example, a particle in a 
double-well potential).
The system can take two stable states which we call state ``0'' and 
state ``1''.
In the situation that the system stays in one of the stable states and
never moves to the other stable state, we say that the system is in a  
``partitioned state''. This can be realized if the temperature is low enough
in comparison with the potential barrier separating the two stable states.
There are two partitioned states corresponding to ``0'' and ``1''.
The system functions as a memory device in one of the partitioned states,
by keeping one bit of information.
Thus we will call the partitioned states as ``m(memory)''-
state which can be ``0'' or ``1''.
 
If the potential has been modified from double-well to single-well, 
the system can 
take only one stable state. Then we say that the system is in 
a ``unified state''.
By the change of the system from a partitioned state to a unified state, 
one bit of information is lost.
We will call the unified state as ``n(neutral)''-state. 

We will study the thermodynamics of 
computation by using two models for a one-bit memory device.

\subsection{Szilard Engine}
 
\hspace*{5mm}
This is an idealized model of a memory device and is helpful to see 
the essence of physics in computational processes.
The engine consists of a molecule, a cylinder, two pistons and a 
partition sheet (Fig. 1).

\begin{figure}[htbp]
\begin{center}
	\includegraphics[width=0.47\linewidth]{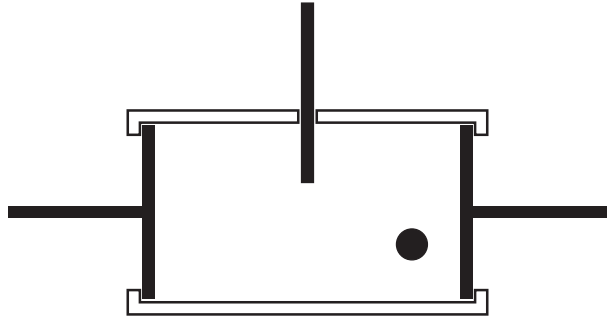}
	\caption{Szilard engine.}
\end{center}
\end{figure} 

The molecule is interacting with a heat bath with the temperature $T$. 
When the partition sheet is inserted, the engine is in one of ``m''-states,
i.e., ``0'' or ``1''.
In state ``0'' ( ``1'' ), the molecule is in the left-hand (right-hand) side. 
When the partition sheet is removed, the engine is in ``n''-state. 

Differences between thermodynamical quantities in ``m''- and ``n''-states
are
\begin{equation}
\begin{array}{llllllll}
{\rm Energy : \quad}      & \Delta E &=& E_{\rm n} - E_{\rm m}  &=& 
0,                   & & \\
{\rm Entropy : \quad}     & \Delta S &=& S_{\rm n} - S_{\rm m}  &=& 
k \ln 2,             & & \\
{\rm Free~ energy : \quad}& \Delta F &=& F_{\rm n} - F_{\rm m}  &=& 
\Delta E - T\Delta S &=& -kT \ln 2.
\label{eq:difference}
\end{array}
\end{equation}

\subsubsection{``Writing process''}

\hspace*{5mm}
``Writing process'' is as follows: 
The initial state is the ``n''-state. 
An agent who 
performs a computation pushes one of the two pistons to the center of the 
cylinder, inserts the partition sheet and returns the piston to the starting 
position. 
In this process the agent does some work on the molecule, which amounts to
\[ 
W_{\rm write}=-\Delta F = kT \ln2.
\]
During this process the entropy $\Delta S$ moves from the system to the
heat bath, i.e., the environment,
corresponding to which the heat generation occurs:
\[
Q_{\rm write}=T \Delta S=kT \ln2.
\]
Evidently, the ``writing process'' is reversible.

\subsubsection{``Reversible-deleting process''}

\hspace*{5mm}
There are two kinds of deleting process. The first one is the inverse of the 
'writing process'. Then the work done by the agent and the heat generation in 
the environment are given by
\begin{equation}
\begin{array}{lllll}
 W_{\rm delete} &=& -W_{\rm write} &=& -kT \ln2,\\
 Q_{\rm delete} &=& -Q_{\rm write} &=& -kT \ln2.
\end{array}
\nonumber
\end{equation}

Therefore the energy necessary for one cycle (``writing''
+``reversible-deleting'') 
is zero and so is the net heat generation:
\[
 W_{\rm rev-cycle} = Q_{\rm rev-cycle} = 0.
\]

However, the above deleting process is possible only when 
the agent knows the content of the memory, i.e., in which state of ``0'' and 
``1'' the system is. This implies that the same information remains in 
some other memory device after the deletion.
 
\subsubsection{``Irreversible-deleting process''}

\hspace*{5mm}
In this process the agent simply pulls out the partition sheet. 
Obviously the agent can do it without any information on the device. 
Both the work and the heat generation are zero in this case:
\[
 W_{\rm irr-delete}=Q_{\rm irr-delete}=0,
\] 
but the entropy is produced in the engine which amounts to
\[
 S_{\rm g}=k \ln2.
\]
The cost of energy for one cycle (``writing''+``irreversible deleting'') and 
the net heat generation are given by
\[
 W_{\rm irr-cycle}=Q_{\rm irr-cycle}=T S_{\rm g}=kT \ln2.
\]

\subsection{A particle in a bistable-monostable potential}

\hspace*{5mm}
This model (see Fig. 2) is mathematically equivalent to a quantum flux 
parametron (QFP) device invented by  Goto\cite{gylh89}. 

\begin{figure}[htbp]
	\begin{center}
		\includegraphics[width=0.55\linewidth]{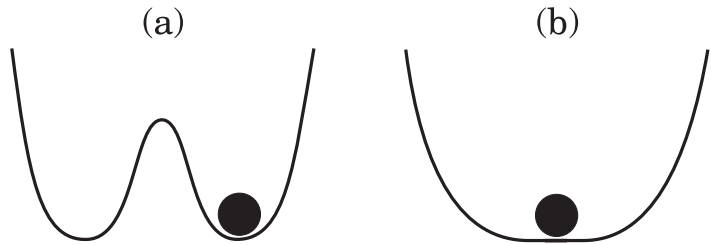}
		\caption{A particle in a bistable-monostable potential. 
(a) ``m''-state,``m''= ``1''. (b) ``n''-state.}
	\end{center}
\end{figure}

\vspace{2mm}
Instead of pushing the pistons in the Szilard engine a 
bias potential is applied in this model. 
The partition sheet is replaced by a 
potential hill at the center.
 It is easily shown that the necessary energy and the net heat for 
one cycle (``writing'' + ``deleting'') are
the same as those in the Szilard engine:
\begin{equation}
\begin{array}{lllllll}
W_{\rm rev-cycle} &=& Q_{\rm rev-cycle} &=& 0,& & \\
W_{\rm irr-cycle} &=& Q_{\rm irr-cycle} &=& T S_{\rm g} &=& kT \ln2.  
\end{array}
\nonumber
\end{equation}
It should be noted that the amount of entropy generated in 
the ``irreversible-deleting process'' is equal to that for 
the Szilard engine. 
The reason is as follows. Entropy
is generated when the height of the central hill is lowered so that 
the particle can jump into the other well by thermal noize. 
At this moment the volume of 
the phase space where the particle can walk around expands twice. 
This means $k \ln2$ of entropy is generated.

We summarize the thermodynamics of computation as follows. Deletion of 
information is physically realized by a ``unifying process'' of 
a ``partitioned state'', which means that the 
barrier (the partition sheet in the Szilard engine, the potential hill 
in the bistable-monostable system) is removed and the two separated 
phase spaces are unified. 
This process is irreversible and accompanied by the entropy production.

\section{Residual entropy}

\hspace*{5mm}
It is widely  believed that a material whose ground state is degenerate has 
$k \ln W_{\rm d}$ of residual entropy at low temperatures, 
where $W_{\rm d}$ is its degeneracy. 
A well-known example is ice whose $W_{\rm d}$ is almost equal to 
$2^N$,  where $N$ is the number of hydrogen atoms. 
The potential energy of a hydrogen atom has two minimum
points between the neighboring two oxygen atoms, which is the origin of the 
degeneracy.
 At low temperatures each hydrogen atom is localized at one of the two 
minimum points. Therefore the ice is in a partitioned state in our terminology.
If the ice really has residual entropy, our memory device also must have 
$k \ln2$ of entropy in ``m''-
state where ``m'' is ``0'' or ``1''. 
However this contradicts the entropy production in the ``irreversible-deleting
 process''.
We believe that the thermodynamics of computation described in the 
previous section really holds. Thus we conclude that the residual entropy does
not exist!

\section{Decrease of entropy and the second law}

\hspace*{5mm}
A ``unifying process'' is an entropy generating process as stated above. 
Then the inverse of the unifying process, that is, a ``partitioning process''
must be an entropy decreasing process. 
The simplest example of this process is to 
insert the partition sheet when the Szilard engine is in the ``n''-state. 
It is evident that the entropy decreases by $k \ln2$.

So far thermodynamics has not taken into account partitioning processes. 
In the following we are discussing how the second law should be modified
when the partitioned state is involved.
 
There are several 'axioms' , 'theorems' or 'lemmas' expressing 
the second law.
We classify them into four 'propositions' and examine them one by one.

\vspace{5mm}
$<$ 1 $>$ Clausius's principle, 
Thomson's principle and Caratheodory's principle.

\vspace{2mm}
These principles hold even when the partitioned states are involved. Namely,
It is not necessary to modify $<$ 1 $>$.

\vspace{5mm}
$<$ 2 $>$ Clausius's inequality:
 For any realizable cycle of a system interacting with an external 
environment, the inequality
\[
       \oint \frac{d^{\prime}Q}{T^{\prime}} \le 0 
\]
holds, where $d^{\prime}Q$ is the heat that flowed into the system 
and  $T^{\prime}$ is the temperature of the heat bath. 
If all of the processes are {\it reversible processes}, 
the equality holds.

\vspace{2mm}
The proposition $<$ 2 $>$ is derived from $<$ 1 $>$ if Carnot's cycle 
is assumed to 
work between heat reservoirs with different temperatures. 
We believe that Carnot's cycle works. 
Then $<$ 2 $>$ holds.
But we must be careful when we use the term {\it reversible process}.
It has been believed that the process A $\rightarrow$ B is reversible 
if B $\rightarrow$ A is reversible. 
However, this is not always true. 
Let B $\rightarrow$ A be a partitioning process. 
Then A $\rightarrow$ B is a unifying process. 
But remember that the partitioning process is reversible while
the unifying process is not.
Therefore the expression '{\it reversible 
process}' in $<$ 2 $>$ must be replaced by '{\it reversible process 
in both directions}'.

\vspace{5mm}
$<$ 3 $>$ For any process A $\rightarrow$ B, the following inequality holds:
\[
     \int \frac{d^{\prime}Q}{T^{\prime}} \le S({\rm B}) - S({\rm A}),
\]
where the equality corresponds to {\it reversible process}.

\vspace{2mm}
The proposition $<$ 3 $>$ works if a cycle made 
of A $\rightarrow$ B and B $\rightarrow$ A satisfies 
the Clausius's inequality, where the process B $\rightarrow$ A is a 
{\it 'reversible process in both directions'}. 
This condition is not satisfied when A $\rightarrow$ B is a partitioning 
process (Szilard's demon). 
Thus it is necessary to replace the expression 'For any process 
A $\rightarrow$ B' by 'If  A $\rightarrow$ B does not include partitioning 
processes'.  
Furthermore, '{\it reversible process}' should be replaced 
by '{\it reversible process in both directions}'.

\vspace{5mm}    
$<$ 4 $>$ In adiabatic systems entropy does not decrease, namely increases or
stays constant.

\vspace{2mm}
The proposition $<$ 4 $>$ is a result obtained by applying $<$ 3 $>$ to 
adiabatic systems.
It is necessary to add a condition that 'if partitioning processes are not
included'.

\section{Concluding remarks}

\hspace*{5mm}
Entropy is a controversial subject. Here we state our understanding.  
Entropy is an objective physical quantity like energy or volume. 
Its relation with information is expressed in eq. (1). 
To record a certain amount of information in a physical system, 
we have to {\it reduce} its entropy by $k \ln2/$bit.

We have introduced the concept of ``partitioned state'', rather intuitively. 
Its unambiguous definition has not yet presented and many questions 
relating with this subject are unanswered.

\vspace{10mm}


\end{document}